\documentclass[preprint,aps,showpacs,superscriptaddress]{revtex4}
\usepackage{graphicx}
\usepackage{dcolumn}
\usepackage{amsfonts, amssymb}
\begin{document}

\title{Faddeev calculations of the $\bar{K}NN$ system with
chirally-motivated $\bar{K}N$ interaction. \\
II. The $K^- pp$ quasi-bound state.}

\author{J. R\'{e}vai}
\affiliation{Wigner Research Center for Physics, RMI,
H-1525 Budapest, P.O.B. 49, Hungary}

\author{N.V. Shevchenko\footnote{Corresponding author:
%phone: +420 266 173 276, fax: +420 220 940 165, e-mail:
shevchenko@ujf.cas.cz}}
\affiliation{Nuclear Physics Institute, 25068 \v{R}e\v{z}, Czech Republic}

\date{\today}

\begin{abstract}
New calculations of the quasi-bound state in the $K^- pp$ system
using Faddeev-type equations in AGS form with coupled $\bar{K}NN$
and $\pi \Sigma N$ channels were performed. A chiral $\bar{K}N$
potential together with phenomenological models of $\bar{K}N$
interaction with one- and two-pole structure of the $\Lambda(1405)$
resonance were used. All three potentials reproduce experimental data
on low-energy $K^- p$ scattering and kaonic hydrogen with the same
level of accuracy. New method of calculating the subthreshold
resonance position and width in a three-body system was proposed
and used together with the direct search of the resonance pole.
We obtained binding energy of the $K^-pp$ quasi-bound state $\sim 32$ MeV
with the chirally motivated and $47 - 54$ MeV with the phenomenological
$\bar{K}N$ potentials. The width is about $50$ MeV for
the two-pole models of the interaction, while the one-pole potential gives
$\sim 65$ MeV width. The question of using an energy dependent potential
in few-body calculations is discussed in detail. It is shown that
``self-consistent'' variational calculations using such a potential are unable
 to produce a reasonable approximation to the exact result.
\end{abstract}

%13.75.Jz: kaon-baryon interactions
%11.80.Gw: Multichannel scattering
%36.10.Gv: Mesonic atoms and molecules, hyperonic atoms and molecules
\maketitle

%%%%%%%%%%%%%%%%%%%%%%%%%%%%%%%%%%%%%%%%%%%%%%%%%%%%%%%%%%%%%%%%
\section{Introduction}
\label{Introduction_sect}

Exotic nuclei and atoms with non-zero strangeness attract attention
from theorists and experimentalists. An assumption, that
a quasi-bound state could exist in the lightest $\bar{K}NN$ system,
appeared quite a long time ago~\cite{Nogami}, the interest to
the topic was renewed after G-matrix calculations of several few-body
antikaonic-nucleus systems~\cite{YA}. Many theoretical calculations
were performed after that,  mostly focusing on the lightest $K^- pp$ system.
All of them used different few-body methods and two-body inputs, as a result
the predicted binding energies and widths of the quasi-bound state differ
substantially. The theoretical results agree only in the fact, that
the quasi-bound state really can exist in the $K^- pp$ system.

The first Faddeev calculations of a quasi-bound state in the $K^- pp$ system
with coupled channels were performed in~\cite{our_KNN_PRL,our_KNN_PRC}
and~\cite{ikedasato1}. The authors of~\cite{ikedasato1} than repeated 
their calculation in~\cite{ikedasato2,ikedasato3} with energy dependent
and independent versions of a chirally motivated $\bar{K}N$ potential.
Other methods with less accurate treatment of the few-body
dynamics were used for investigation of the $K^- pp$ system after the
Faddeev calculations, in particular, one-channel variational
approaches~\cite{vaizeyap1,vaizeyap2,evrei} and some others.

The first experimental evidence of a quasi-bound state observation in
$K^- pp$ occurred in the FINUDA experiment~\cite{FINUDA} at the DA$\Phi$NE
$e^+ e^−$ collider.
New analyses of old experiments, such as OBELIX~\cite{OBELIX} at CERN and
DISTO~\cite{DISTO} at SATURNE also claim the observation of the state.
The experimental results like the theoretical ones differ from each other,
moreover, their binding energies and widths are far from all theoretical
predictions. Since the
question of the possible existence of the quasi-bound
state in $K^- pp$ system is still highly actual, new experiments are
being planned and performed by HADES~\cite{HADES} and LEPS~\cite{LEPS}
Collaborations, in J-PARC E15~\cite{J-PARC_E15} and E27~\cite{J-PARC_E27}
experiments.

The theoretical works differ not only in methods of treatment of the few-body
system and models of two-body interactions, but also in accuracy of 
reproducing experimental data on $K^- p$ scattering and kaonic hydrogen
by the $\bar{K}N$ potentials. The $\bar{K}N$ interaction, which is the most
important for the $\bar{K}NN$ system, is usually described either by pure
phenomenological or by chirally motivated potentials. In particular, our
previous calculations of
the quasi-bound state in the $\bar{K}NN$ system~\cite{our_KNN_PRL,our_KNN_PRC}
were performed with an early phenomenological model of $\bar{K}N$ interaction.
In a series of subsequent works devoted to the $K^- d$
system~\cite{ourPRC_isobreak,my_Kd,my_Kd_sdvig}
more accurate phenomenological models of $\bar{K}N$ interaction were constructed
and used. The potentials with coupled $\bar{K}N$ and $\pi \Sigma$ channels
reproduce low-energy data on $K^- p$ scattering and kaonic hydrogen and have
a one-pole and a two-pole structure of the $\Lambda(1405)$ resonance, while
the potential from~\cite{our_KNN_PRL,our_KNN_PRC} has only one pole.
In addition, a chirally motivated $\bar{K}N - \pi \Sigma - \pi \Lambda$ potential,
which reproduces the experimental data with the same accuracy as the
phenomenological ones, was constructed~\cite{I}. We used all three
potentials in the new calculations of the $K^- pp$ quasi-bound state and
thereby investigated the dependence of the three-body results on the models
of the $\bar{K}N$ interaction.

Faddeev equations in AGS form and the two-body input, which were used in
the calculations, are shortly described in the next section together with
the methods of evaluating the binding energy and width of a quasi-bound state.
Two methods were used: the direct search of the pole position
with contour rotation (described in Subsection~\ref{contour.sect}) and a new
method of finding a subthreshold resonance (\ref{Det_method.sect}).
Our exact results are presented and discussed in Section~\ref{Results.sect},
which also contains our results of approximate calculations, performed
additionally for comparison of our characteristics of the $K^- pp$ quasi-bound
states with those, obtained by other authors. Section~\ref{Zdep.sect} is devoted
to the question of using of
an energy dependent potential in few-body equations. Series of
additional calculations were performed to investigate the applicability of
the ``self-consistent'' method of obtaining an ``averaged $\bar{K}N$ energy
in $\bar{K}NN$ system'' used in the variational calculations. Our conclusions 
are drawn in the last section.

%%%%%%%%%%%%%%%%%%%%%%%%%%%%%%%%%%%%%%%%%%%%%%%%%%%%%%%%%%%%%%%%%%%%%%%%%%%%%%%
\section{Method}
\label{Method.sect}

The Faddeev equations in Alt-Grassberger-Sandhas form for
the three-body system with coupled $\bar{K}NN$ and $\pi \Sigma N$ channels are
described in detail in our previous papers~\cite{our_KNN_PRC,my_Kd}.
The equations are written for separable two-body potentials, they are properly
antisymmetrized. The homogeneous system of 10 integral equations schematically
can be written as
\begin{equation}
\label{AGS_imp}
 X_{i}(p) = \int_{0}^{\infty} Z_{ij}(p,p';z) \,
 \tau_{j}\left(z - \frac{p'^2}{2 \mu_j}\right) X_{j}(p') dp',
\end{equation}
where $X_i$ is an unknown function and $\tau_j$ is the energy dependent
part of a two-body $T$-matrix describing the interaction of the particles $(ik)$
(as is usual for Faddeev equations, $i \ne j \ne k$ is assumed), corresponding
to a separable potential
\begin{equation}
\label{Vsep}
 V(k,k') = g(k) \, \lambda \, g(k') 
  \quad \longrightarrow \quad T(k,k';z^{(2)}) = g(k) \, \tau(z^{(2)}) \, g(k').
\end{equation}
The energy $z^{(2)}$ in~Eq.(\ref{Vsep}) is an energy of
a two-body system, while $z$ in~Eq.(\ref{AGS_imp}) is the three-body energy.
Momentum $k$ in~Eq.(\ref{Vsep}) describes motion of a pair of particles, while
$p$ in~Eq.(\ref{AGS_imp}) is a momentum
of relative motion of a particle in respect to a pair. All additional indices
and summations in the Eqs.~(\ref{AGS_imp},\ref{Vsep}) are omitted.

All our potentials are $s$-wave isospin dependent ones.
We used three different models describing the $\bar{K}N$ interaction, which
is the most important for the $\bar{K}NN$ system.
Two phenomenological potentials with one-pole $V^{1,SIDD}_{\bar{K}N-\pi \Sigma}$
and two-pole $V^{2,SIDD}_{\bar{K}N-\pi \Sigma}$ structure of the $\Lambda(1405)$
resonance from~\cite{my_Kd_sdvig} describe directly coupled $\bar{K}N$ and
$\pi \Sigma$ channels, while the $\pi \Lambda$ channel is taken into account
effectively through complex value of one of the strength parameters.
Our chirally motivated $\bar{K}N - \pi \Sigma - \pi \Lambda$ potential
with three coupled channels is described in I~\cite{I}. All three models
are equally good in reproducing experimental data
on low-energy $K^- p$ scattering and kaonic hydrogen. In particular, they
reproduce elastic and inelastic $K^- p$ cross-sections and threshold branching
ratio $\gamma$. The remaining threshold branching ratios $R_c$ and $R_n$ are
reproduced by the $V^{Chiral}_{\bar{K}N}$ directly, while auxiliary $R_{\pi \Sigma}$
constructed from $R_c$ and $R_n$ is reproduced by the phenomenological
potentials instead due to absence of the directly coupled $\pi \Lambda$ channel.
All three models of $\bar{K}N$ interaction give values for $1s$ level shift and
width of kaonic hydrogen, which are in agreement with the most recent experimental
results from SIDDHARTA Collaboration~\cite{SIDDHARTA}.

The remaining two-body potentials, used the three-body calculation,
are described in~\cite{my_Kd}. The two-term TSA-B $NN$ potential reproduces
phase shifts of Argonne V18 potential, therefore, is repulsive at short distances.
It also gives proper $NN$ scattering length, effective range and binding energy
of deuteron. As for the model of $\Sigma N$ interaction,
we used the spin independent version of the exact optical potential
corresponding to the model with coupled $\Sigma N$ and $\Lambda N$ channels.
The two-channel $\Sigma N -\Lambda N$ potential reproduces low-energy experimental
$\Sigma N$ and $\Lambda N$ cross-sections, the exact optical $\Sigma N (-\Lambda N)$
potential, which we used, has exactly the same elastic $\Sigma N$ amplitude
as the two-channel model.

The position $z_0$ of a quasi-bound state in the three-body problem is usually
defined as $\lambda(z_0) = 1$, where $\lambda(z_0)$ is an eigenvalue of the
kernel of Faddeev equations. In practice this amounts to
solving the equation ${\rm Det}(z_0) = 0$, where ${\rm Det}(z)$ is
the determinant of the linear system, obtained after discretization of
the integral equations Eq.~(\ref{AGS_imp}). We used two different types of
discretization. One is based on the method of quadratures, another one uses
a cubic spline expansion. All our results obtained with these two methods
are equal, coinciding in $3-4$ significant digits. We also used two methods
of searching the complex pole position in a three-body system.

%%%%%%%%%%%%%%%%%%%%%%%%%%%%%%%%%%%%%%%%%%%%%%%%%%%%%%%%%%%%%%%%%%%%%%%%%%%%%%%
\subsection{Direct pole search with contour rotation}
\label{contour.sect}

Applying the coupled-channel Faddeev formalism for the search of $K^- pp$
quasi-bound state at a complex energy one has to be aware of
the specific requirements of finding a three-body resonance pole on the
proper Riemann sheet. The problem of proper analytic continuation of the
momentum space Faddeev equations from the physical energy sheet has an
extended literature, see e.g.~\cite{Contur1,Contur2,Contur3,Contur4,Contur5our}.
It has been established, that the correct analytic continuation to the
closest unphysical energy sheet can be achieved by moving the momentum
integration into the complex plane. Different integration contours were
proposed in the literature. We chose the one suggested in~\cite{Contur5our},
which is a ray in the fourth quadrant of the complex plane. Along the ray
the momentum variable $p'$ from Eq.(\ref{AGS_imp}) must satisfy the condition
$|{\rm Arg}(p')| > |{\rm Arg}(z_0)|/2$. Deformation of the integration contour
in this way ensures, that the complex solution of the equation will be found
on the correct energy sheet.

%%%%%%%%%%%%%%%%%%%%%%%%%%%%%%%%%%%%%%%%%%%%%%%%%%%%%%%%%%%%%%%%%%%%%%%%%%%%%%%
\subsection{$1/|{\rm Det}|^2$ method}
\label{Det_method.sect}

\begin{figure*}
\centering
\includegraphics[width=0.80\textwidth]{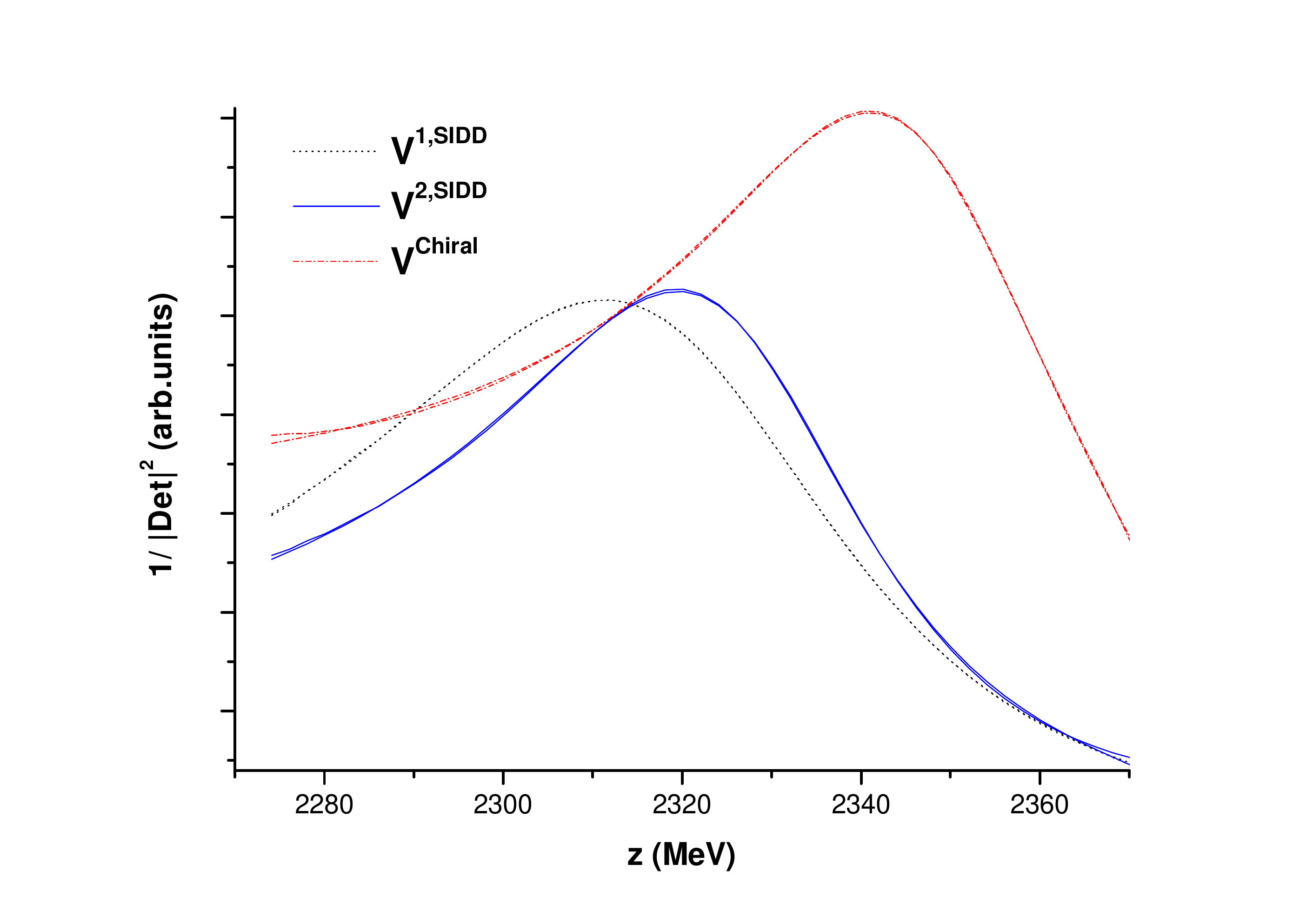}
\caption{Function $1/|{\rm Det}(z)|^2$ calculated with the
one-pole $V^{1,SIDD}_{\bar{K}N-\pi \Sigma}$ (black dotted line),
two-pole $V^{2,SIDD}_{\bar{K}N-\pi \Sigma}$ phenomenological
potentials (blue solid line) and chirally motivated
$V_{\bar{K}N - \pi \Sigma - \pi \Lambda}^{\rm Chiral}$ potential
(red dash-dotted line). Breit-Wigner fits for all three functions
are almost indistinguishable from the original lines.
\label{1Det.fig}}
\end{figure*}

We found also another way to locate the $K^- pp$ quasi-bound state, which avoids
integration in the complex plane. Since the function ${\rm Det}(z)$ has a zero at
the quasi-bound state position $z_0$, the function $1/{\rm Det}(z)$, which
enters all amplitudes through the inverse Faddeev matrix, can be written as
\begin{equation}
\label{1Det}
 \frac{1}{{\rm Det}(z)} = \frac{d(z)}{z - z_0}.
\end{equation}
When energy $z$ in Eq.(\ref{1Det}) is taken on the real axes, the function
$1/|{\rm Det}(z)|^2$ has a Breit-Wigner form with $d(z)$ accounting for a
background. Therefore, calculating
${\rm Det}(z)$ for real energies $z$, for which the integration on momentum $p'$
in Eq.(\ref{AGS_imp}) can be kept on the real axis, and fitting $1/|{\rm Det}(z)|^2$
function to a Breit-Wigner curve we can get information on the resonance position
and width. Obviously, this procedure works only if the resonance is well pronounced,
i.e. isolated and not too wide. Fortunately, our $K^- pp$ quasi-bound state is
of this type, as can be seen on Fig.\ref{1Det.fig}, where we show the results
of applying this method. The function $1/|{\rm Det}(z)|^2$, calculated on the real
energy axis using all three our $\bar{K}N$ potentials, is demonstrated.
The Breit-Wigner fits to the curves are also plotted, however, they
are almost indistinguishable from the original lines. The fits
were done with a background function, which is quadratic in energy. It is
remarkable, that the shape of the $1/|{\rm Det}(z)|^2$ function on the real energy
axis does not depend on the actual method of discretization
of the integral equations. We obtained almost strictly coinciding Breit-Wigner
parameters from quadratures and cubic spline expansion discretizations
in spite of the fact, that the determinants themselves were strongly different.
%%%%%%%%%%%%%%%%%%%%%%%%%%%%%%%%%%%%%%%%%%%%%%%%%%%%%%%%%%%%%%%%%%%%%%%%%%
%%%%%%%%%%%%%%%%%%%%%%%%%%%%%%%%%%%%%%%%%%%%%%%%%%%%%%%%%%%%%%%%%%%%%%%%%%

Since complex root finding is a difficult task, the Breit-Wigner values can
serve as a good starting point for it. On the other hand, it is
a good test of the directly found pole position, which is free from
the possible uncertainty of the proper choice of the Riemann sheet.
%----------------------------------------------------------------
\begin{center}
\begin{table}
\caption{Pole positions $z_{K^-pp}$ (in MeV, the real part is measured from
the $\bar{K}NN$ threshold) of the quasi-bound state in the $K^- pp$ system:
the results of the direct pole search and of the Breit-Wigner fit of the
$1/|{\rm Det(z)}|^2$ function at real energy axis. The AGS calculations
performed with the one-pole $V^{1,SIDD}_{\bar{K}N-\pi \Sigma}$,
two-pole $V^{2,SIDD}_{\bar{K}N-\pi \Sigma}$ phenomenological
potentials from~\cite{my_Kd_sdvig} and the chirally-motivated
$V^{Chiral}_{\bar{K}N - \pi \Sigma - \pi \Lambda}$ potential from~\cite{I}
are demonstrated.}
\label{Pole1Det.tab}
\begin{center}
\begin{tabular}{ccc}
\hline \noalign{\smallskip}
  & \qquad Direct pole search & \qquad BW fit of $1/|{\rm Det(z)}|^2$ \\
\hline \noalign{\smallskip}
 with $V^{1,SIDD}_{\bar{K}N}$, \cite{my_Kd_sdvig} 
           & $-53.3 -  i \, 32.4$ & $-54.0 -  i \, 33.3$ \\
 with $V^{2,SIDD}_{\bar{K}N}$, \cite{my_Kd_sdvig}  
           & $-47.4 -  i \, 24.9$ &  $-46.2 -  i \, 25.9$ \\
 with $V_{\bar{K}N}^{\rm Chiral}$, \cite{I}  
           & $-32.2 -  i \, 24.3$ & $-30.3 -  i \, 23.3$ \\
           \noalign{\smallskip} \hline
\end{tabular}
\end{center}
\end{table}
\end{center}
%----------------------------------------------------------------

%%%%%%%%%%%%%%%%%%%%%%%%%%%%%%%%%%%%%%%%%%%%%%%%%%%%%%%%%%%%%%%%%%%%%%%%%%%%%%%
\section{Results}
\label{Results.sect}

Pole positions of the $K^- pp$ quasi-bound state, obtained from the direct
search in the complex plane and from the Breit-Wigner fit with three 
our $\bar{K}N$ potentials are shown in Table~\ref{Pole1Det.tab}.
The one- $V^{1,SIDD}_{\bar{K}N}$ and two-pole
$V^{2,SIDD}_{\bar{K}N}$ phenomenological interaction models
from \cite{my_Kd_sdvig} together with the chirally motivated potential
$V_{\bar{K}N}^{\rm Chiral}$ from \cite{I} were used. It is seen,
that the results obtained using the two methods of pole position search
are quite close, indicating, that the methods supplement each other.
%----------------------------------------------------------------
\begin{center}
\begin{table}
\caption{Binding energy $B_{K^- pp}$ (MeV) and width $\Gamma_{K^- pp}$ (MeV)
of  the quasi-bound state in the $K^- pp$ system. The results obtained
from AGS calculation with the one-pole $V^{1,SIDD}_{\bar{K}N-\pi \Sigma}$,
two-pole $V^{2,SIDD}_{\bar{K}N-\pi \Sigma}$ phenomenological
potentials from~\cite{my_Kd_sdvig} and the chirally-motivated
$V^{Chiral}_{\bar{K}N - \pi \Sigma - \pi \Lambda}$ potential from~\cite{I}
are demonstrated. Other theoretical results of Faddeev
\cite{our_KNN_PRC,ikedasato3} and variational \cite{vaizeyap2,evrei}
calculations are also shown.}
\label{BGam.tab}
\begin{center}
\begin{tabular}{ccc}
\hline \noalign{\smallskip}
  & $B_{K^- pp}$ & $\Gamma_{K^- pp}$ \\
\hline \noalign{\smallskip}
\underline{Present AGS:}  & & \\
 with $V^{1,SIDD}_{\bar{K}N}$, \cite{my_Kd_sdvig}  & 53.3 & 64.8 \\
 with $V^{2,SIDD}_{\bar{K}N}$, \cite{my_Kd_sdvig}  & 47.4 & 49.8 \\
 with $V_{\bar{K}N}^{\rm Chiral}$, \cite{I}  & 32.2 & 48.6 \\
 \noalign{\smallskip} \hline \noalign{\smallskip}
\underline{Previous AGS:}  & & \\
 SGMR~\cite{our_KNN_PRC}  & 55.1 & 100.2 \\
 IKS~\cite{ikedasato3} with $V^{E-indep}_{\bar{K}N}$ & 44 - 58 & 34 - 40 \\
 IKS~\cite{ikedasato3} with $V^{E-dep}_{\bar{K}N}$ & 9 - 16  & 34 - 46 \\
                          & 67 - 89  & 244 - 320 \\
\noalign{\smallskip} \hline \noalign{\smallskip}
\underline{Variational:} & & \\
 DHW~\cite{vaizeyap2} & 17 - 23 & 40 - 70 \\
 BGL~\cite{evrei}     & 15.7    & 41.2  \\
\noalign{\smallskip} \hline
\end{tabular}
\end{center}
\end{table}
\end{center}
%----------------------------------------------------------------

The most striking feature of the results, shown in the table, is the large
difference between the binding energies of the quasi-bound states obtained
from the phenomenological, especially the one-pole version, and the chirally
motivated $\bar{K}N$ potentials. This probably is due to the energy dependence
of chirally motivated model of the interaction. 
The available experimental data, to which all three potentials
were fitted with approximately equal accuracy are close to the $\bar{K}N$
threshold. While the phenomenological models of the $\bar{K}N$ interaction
are unchanged, when the $K^- pp$ quasi-bound state is calculated, the energy
dependence of the chirally motivated potential reduces the attraction for
the lower energies in the $\bar{K}NN$ quasi-bound state region, thus producing
states with less binding.

It is not quite clear, why the widths of the two-pole models of $\bar{K}N$
interaction are almost coinciding, while the one-pole $V^{1,SIDD}_{\bar{K}N}$
potential gives much larger width. The difference might be connected with the
different pole stucture of the corresponding $\bar{K}N$ interaction models:
while the highest poles of the two-pole $V^{2,SIDD}_{\bar{K}N}$ and chirally
motivated $V_{\bar{K}N}^{\rm Chiral}$ potentials lie close to each other,
the pole position of the one-pole phenomenological model is much closer to
the $\bar{K}N$ threshold (see Table 2 of \cite{my_Kd_sdvig} and Eq.(12) of
\cite{I}).

Such a large difference between the ``phenomenological'' and ``chiral'' results
is opposite to the results of I~\cite{I}, where low-energy $K^- d$ scattering
and $1s$ level shift and width of kaonic deuterium were calculated using the same
equations (surely, inhomogeneous ones with correspondingly changed quantum numbers)
and input. In that case the three-body observables obtained with the three
$\bar{K}N$ potentials turned out to be very close each to other. It can be
due to the fact, that those three-body values were calculated near the
$\bar{K}NN$ threshold while the $K^- pp$ pole positions are far below it.

Our three binding energy $B_{K^- pp}$ and width $\Gamma_{K^- pp}$ values of 
the $K^- pp$ quasi-bound state are compared in Table~\ref{BGam.tab} with other
theoretical results. In particular, the results obtained in our previous
Faddeev calculation~\cite{our_KNN_PRC}, the most recent results of alternative
calculation using the same equations~\cite{ikedasato3} with several
chirally motivated $\bar{K}N$ potentials are shown together with
two variational results~\cite{vaizeyap2,evrei}.
The new result with one-pole $V^{1,SIDD}_{\bar{K}N}$ potential has binding
energy, which is very close to our previous one~\cite{our_KNN_PRC}, which used
the same model of $\bar{K}N$ interaction. The difference in widths could
be explained by low accuracy of the older $\bar{K}N$ potential.

Coupled-channel AGS equations were solved in~\cite{ikedasato3}
with chirally motivated energy dependent and independent $\bar{K}N$
potentials. Therefore, in principle, their calculation with the energy
dependent version of the $\bar{K}N$ potential $V^{E-dep}_{\bar{K}N}$
should give a result, which is close to ours with chirally motivated
model of interaction $V_{\bar{K}N}^{\rm Chiral}$. It is seen, however,
that only their width is comparable to our $\Gamma_{K^- pp}$, while
the binding energy obtained in~\cite{ikedasato3} is much smaller than ours.
The reason of the difference is, probably, an approximation used
in the chirally motivated models used in~\cite{ikedasato2,ikedasato3}.
Namely, the energy-dependent square root factors, responsible for the
correct normalization of the amplitudes, are replaced by constant masses.
This can be reasonable for the highest $\bar{K}N$ channel, however,
it is certainly a poor approximation for the lower lying $\pi \Sigma$ and
$\pi \Lambda$ channels. We checked the role of this approximation in
the AGS calculations, the obtained pole position, corresponding to the
quasi-bound state,
\begin{equation}
 z_{K^- pp}^{Const.norm} = -25.0  - i \, 23.4 \; {\rm MeV}
\end{equation}
really has much smaller binding energy than the original one, see
Table~\ref{BGam.tab}. The remaining difference between the results could be
explained by the higher accuracy of reproducing experimental $K^- p$ data
by our chirally motivated $\bar{K}N$ potential.

We did not find the second pole in the $K^- pp$ system reported
in~\cite{ikedasato3} in the corresponding region for either of the
three $\bar{K}N$ potentials.

There are a few problematic points in~\cite{vaizeyap2,evrei}, too.
First of all, the variational calculation was performed solely in the
$\bar{K}NN$ sector, therefore the absorption into the $\pi \Sigma N$ and
$\pi \Lambda N$ channels should be taken into account through the imaginary
part of an optical or complex potential. We checked an accuracy of
use of the exact optical $\bar{K}N$ potential, which gives exactly the same
elastic $\bar{K}N$ amplitude as the original potential with coupled channels,
and performed one-channel AGS calculations for three our $\bar{K}N$ potentials.
The ``exact optical'' pole positions
\begin{eqnarray}
 z_{K^- pp}^{1,SIDD,Opt} &=& -54.2  - i \, 30.5 \; {\rm MeV} \\
 z_{K^- pp}^{2,SIDD,Opt} &=& -47.4  - i \, 23.0 \; {\rm MeV} \\
 z_{K^- pp}^{Chiral,Opt} &=& -32.9  - i \, 24.4 \; {\rm MeV}
\end{eqnarray}
differ only slightly from the full coupled-channel results from 
Table~\ref{Pole1Det.tab}. Therefore, the one-channel Faddeev calculation
with exact optical potential could be quite satisfactory approximation
to the full calculation with coupled channels. The authors of variational
calculations~\cite{vaizeyap2,evrei} used a one-channel $\bar{K}N$ potential,
derived from a chirally motivated model of interaction with many couped channels.
However, the potential cannot be called ``the exact optical'' since Gauss
form-factors were additionally introduced into the potential. It is not quite
clear, how this one-channel potential is connected to the original one and
whether it still reproduces some experimental $\bar{K}N$ data.

Moreover, the position of the quasi-bound state was determined
in~\cite{vaizeyap2,evrei} only from the real part of this $\bar{K}N$ potential,
as a real bound state, while the width was estimated as the expectation value of
the imaginary part of the potential. This, essentially perturbative treatment of
the inelasticity might be justified for quite narrow resonances, but the
quasi-bound state under consideration is certainly not of this type.

Another serious problem of the variational calculations is their method of
treatment of the energy dependence of the $\bar{K}N$ potential in the
few-body calculations. Our results already show that the energy dependence
of the chirally motivated model of $\bar{K} N$ interaction has a crucial effect
on the $K^- pp$ quasi-bound state position. Therefore, the question deserves
special attention.

%%%%%%%%%%%%%%%%%%%%%%%%%%%%%%%%%%%%%%%%%%%%%
%\section{Treatment of energy dependence}
\section{Energy dependent $\bar{K}N$ potential in few-body calculations}
\label{Zdep.sect}

The basic problem is, that two-body energy is not a well defined quantity in
more-than-two-body systems. Therefore some special effort is needed if energy
dependent interactions are used in few-body calculations. Fortunately, momentum
space Faddeev equations provide a framework, offering an exact treatment of
this problem. It is seen from Eqs.(\ref{AGS_imp}) and (\ref{Vsep}) that the argument
of the energy dependent part $\tau_j$ of the two-body $T$-matrix
\begin{equation}
\label{z2}
z^{(2)}_j \equiv z - \frac{p'^2}{2 \mu_j}
\end{equation}
comes from embedding the two-body $T$-operator in three-body space. The
kinetic energy of the third, non-interacting particle with momentum
$p'$ in Eq.(\ref{z2}) is extracted from the three-body energy $z$. This is
the so called spectator mechanism, through which two-body dynamics enters
the three-body problem in Faddeev approach. The representation of a few-body
system as an auxiliary subsystem of two interacting particles and a third,
non-interacting spectator allows an exact definition of the two-particle energy
at which the two-body $T$-matrix has to be evaluated.
It is seen from Eq.(\ref{AGS_imp}), that since the spectator momentum $p'$
is an integration variable, a range of two-particle energies 
$z^{(2)}_j \subset[z,- \infty)$ enters the three-body calculation.

Our results, shown in the previous section, were obtained in this way, allowing
for a ``dynamical'' dependence on two-body energies. However, we made an
exception: since the coupling constants depending on $z^{(2)}_j$ obviously
become non-physical for $p' \to \infty$, we ``froze'' the energy dependence
at the $\pi \Lambda$ threshold, where
${\rm Re} \, z^{(2)}_j = m_{\pi} +  m_{\Lambda}$,
and kept these values, when $p'$ was further increased under the integral.
We think, that this procedure is non-contradictory to the spirit of chiral
interactions, whose energy dependence is probably meant in a certain region
near the channel thresholds, and definitely not below the lowest open channel.
We checked, that this freezing does not change the pole position,
situated far above the $\pi \Lambda$ threshold.

When chiral interactions are used in non-Faddeev few-body calculations, their
energy dependence has to be accounted for. To treat this problem in coordinate
space variational calculations, which can be performed only with fixed-energy
two-body interactions, a method was invented  in~\cite{vaizeyap1} and
then used in~\cite{vaizeyap2,evrei}. It is based on the assumption, that a definite
two-particle energy $z_{\bar{K}N}$ exists, for which the chiral interaction fixed
at this value produces the same $K^- pp$ quasi-bound state, as the energy
dependent one. A ``self-consistent'' iterative procedure was suggested, according
to which the expectation value of a certain operator, calculated with the trial
wave function and called ``average $\bar{K}N$ energy in $\bar{K}NN$ system'',
should give $z_{\bar{K}N}$. Apart from the fact that the quantity, with respect
to which self-consistency is sought, is not free of some amount of arbitrariness,
no single hint concerning the applicability or accuracy of this method is given.

\begin{figure*}
\centering
\includegraphics[width=0.8\textwidth]{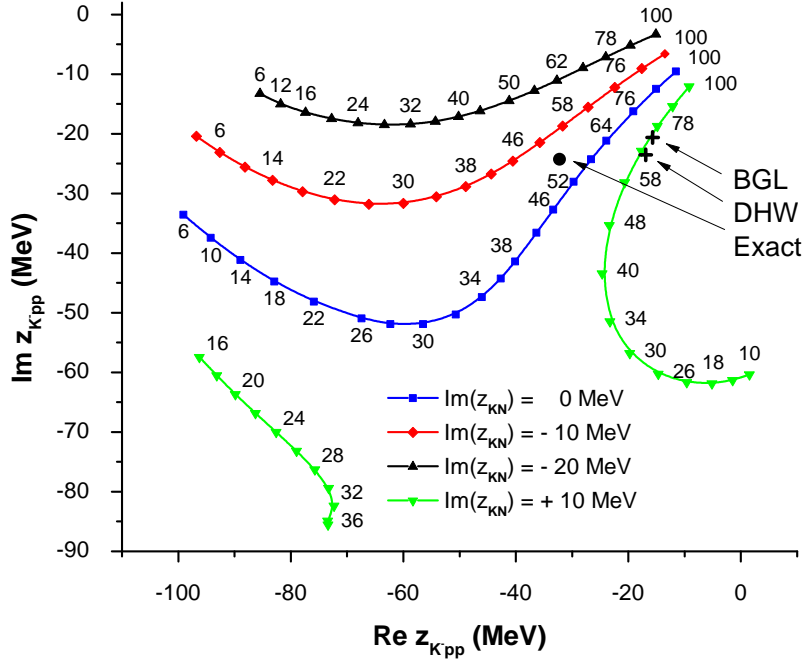}
\caption{Series of the quasi-bound system pole positions in the $K^- pp$
system calculated with fixed energy of the chirally motivated $\bar{K}N$
potential. Each line contains results obtained with
${\rm Im} \, z_{\bar{K}N} = -20$ MeV (black triangles up),
${\rm Im} \, z_{\bar{K}N} = -10$ MeV (red diamonds),
${\rm Im} \, z_{\bar{K}N} = 0$ MeV (blue squares) and
${\rm Im} \, z_{\bar{K}N} = +10$ MeV (green triangles down)
with $|{\rm Re} \, z_{\bar{K}N}|$, changing up to $100$ MeV (numbers near
the lines). Exact result of Faddeev calculation with coupled channels
(black circle) and two results of variational calculations (crosses)
are also shown: BGL~\cite{evrei} and type I result with HNJH $\bar{K}N$
potential DHW~\cite{vaizeyap2}.
\label{EKN_fixed.fig}}
\end{figure*}

We decided to investigate the effect of fixing the energy of the $\bar{K}N$ coupling
functions on the $\bar{K}NN$ quasi-bound state position. Our results are shown in
Fig.\ref{EKN_fixed.fig}, where the $K^- pp$ quasi bound state pole trajectories,
calculated with a fixed $z_{\bar{K}N}$ in the couplings of our chirally motivated
$\bar{K}N$ potential are plotted. The curves correspond to changing
${\rm Re} \, z_{\bar{K}N}$ up to $-100$ MeV, keeping ${\rm Im} \, z_{\bar{K}N}$ fixed.
$|{\rm Re} \, z_{\bar{K}N}|$ values are marked on the plot near the corresponding
points. The curve with ${\rm Im} \, z_{\bar{K}N} = 0$
corresponds to the line, on which, according to~\cite{vaizeyap2,evrei},
a self-consistent procedure can find the correct quasi-bound state position.
We see, that this claim is unjustified, especially for the self-consistent values
of ${\rm Re} \, z_{\bar{K}N}$ found in~\cite{vaizeyap2,evrei}:
\begin{eqnarray}
\label{zKNvaize}
 {\rm Re} \, z_{\bar{K}N}^{DHW} &=& -39 \; {\rm MeV} \\
\label{zKNevrei}
 {\rm Re} \, z_{\bar{K}N}^{BGL} &=& -43 \; {\rm MeV.}
\end{eqnarray}
%%%%%%%%%%%%%%%%%%%%%%%%%%%%%%%%%%%%%%%%%%%%%%%%%%%%%%%%%%%%%%%%%%%%%%%%%%
%%%%%%%%%%%%%%%%%%%%%%%%%%%%%%%%%%%%%%%%%%%%%%%%%%%%%%%%%%%%%%%%%%%%%%%%%%

The curves with non-zero ${\rm Im} \, z_{\bar{K}N}$ show an interesting ``inverse''
behavior: with increasing $|{\rm Im} \, z_{\bar{K}N}|$ the quasi-bound state becomes
narrower. This is due to the fact, that $z_{\bar{K}N}$ enters the diagonal couplings
of the chiral potential with negative sign. Thus increasing of the negative
${\rm Im} \, z_{\bar{K}N}$ by absolute value corresponds not to increasing of the
absorption due to the open $\pi \Sigma N$ channel, but to its reduction.

From Fig.\ref{EKN_fixed.fig} one can also conclude, that the imaginary part of
$z_{\bar{K}N}$ influences not only the width of the quasi-bound state, but its
position, too. Otherwise, the points on the curves, corresponding to equal
${\rm Re} \, z_{\bar{K}N}$ values, should lie strictly below each other. The deviations
from this pattern are apparent, especially towards the ends of trajectories.

If one asks the question, whether a similar curve can be found, which contains
the exact pole, the answer is ``yes'', with
${\rm Im} \, z_{\bar{K}N} \simeq -7$ MeV, while the pole is found at
${\rm Re} \, z_{\bar{K}N} \simeq -58$ MeV. This value of $z_{\bar{K}N}$ is rather
far from the ``self-consistent'' ones, Eqs.(\ref{zKNvaize},\ref{zKNevrei}).
Moreover, it is also possible to find $z_{\bar{K}N}$ values, which yield the
quasi-bound state positions, found in~\cite{vaizeyap2,evrei}, also shown on
Fig.\ref{EKN_fixed.fig}. These values, however, have positive imaginary parts,
which is hard to interpret. It is interesting to note that for positive fixed
${\rm Im} \, z_{\bar{K}N}$ above a certain (critical) value the trajectories
have two branches, as it is seen on the example with 
${\rm Im} \, z_{\bar{K}N} = +10$ MeV. As a consequence, some values of 
$z_{\bar{K}N}$ allow for two poles in the considered region of the energy plane.

In general, it looks like for any quasi-bound state
location one can find a complex $z_{\bar{K}N}$, which in a coupled-channels
Faddeev calculation leads to this pole position. However, even if we know, that
there exists a $z_{\bar{K}N}$ giving the correct quasi-bound state pole, it is
absolutely not clear, whether an operator can be defined, whose expectation value
would give this $z_{\bar{K}N}$, at least approximately. Without such an operator
no self-consistent scheme can be constructed to treat the energy dependence of
the interaction.

\section{Conclusions}
\label{conclusions.sect}

We calculated $\bar{K}NN$ quasi-bound state positions for the two phenomenological
and the chirally motivated models of the $\bar{K}N$ interaction, which all describe
the available experimental $K^- p$ data equally well. We found, that the quasi-bound
states resulting from the phenomenological potentials lie about $15-20$ MeV deeper,
than those of the chirally motivated one. In our opinion, this is due to the energy
dependence of the chiral interaction, leading to less attraction for lower energies.
We obtained binding energy $\sim 32$ MeV for the chirally motivated
and $47 - 54$ MeV for the phenomenological $\bar{K}N$ potentials. The width is about
$50$ MeV was obtained with the two-pole models of the interaction, while the one-pole
potential gives $\sim 65$ MeV.

We proposed a new $1/|{\rm Det}(z)|^2$ method of finding mass and width of
a subthreshold resonance and demonstrated its efficiency.

We discussed in some detail, how energy dependence of the two-body interaction can
be accounted for in few-body calculations. It was shown, that momentum space Faddeev
integral equations allow an exact treatment of this energy dependence. On the contrary,
coordinate space variational methods can use only  energy independent interactions,
therefore we performed a series of calculations with differently fixed two-particle
$\bar{K}N$ energies $z_{\bar{K}N}$ in the couplings of the chirally motivated
interaction. Our conclusion is, that the method used in~\cite{vaizeyap2,evrei}
is unable to define an ``averaged'' $z_{\bar{K}N}$, for which the fixed-energy
chirally motivated interaction, even in a correct three-body calculation, can
yield a $K^- pp$ quasi-bound state position with any relation to the exact one. 

First, our calculations show, that a real $z_{\bar{K}N}$ has absolutely no chance
to reproduce or reasonably approximate the exact quasi-bound state position, even
with correct treatment of the imaginary part of the interaction, unlike
in~\cite{vaizeyap2,evrei}. Second, the way, how the ``self-consistent''
value of (generally complex) $z_{\bar{K}N}$ is defined in the papers does not seem
to guarantee, that the correct value will be reached or at least approximated.
In view of the above considerations, the results of~\cite{vaizeyap2,evrei}
can be considered as rough estimates of what a really energy dependent $\bar{K}N$
interaction will produce in the $K^- pp$ system. Similarly, the four-body results
of~\cite{evrei} would hardly survive a comparison with an exact four-body
calculation, which, however, still has to be done.

\vspace{5mm}

\noindent
{\bf Acknowledgments.}
The work was supported by the Czech GACR grant P203/12/2126 and the Hungarian
OTKA grant 109462.

\end{document}